\documentclass[%
reprint,
superscriptaddress,
 amsmath,amssymb,
 prd,onecolumn
]{revtex4-2}

\usepackage{graphicx}
\usepackage{color}
\usepackage{amssymb}
\usepackage{amsmath}
\usepackage{array}
\usepackage{mathtools}
\usepackage{float}
\usepackage{tabularx}
\usepackage{adjustbox}
\usepackage{hyperref}
\usepackage{url}
\usepackage{natbib}
\usepackage[normalem]{ulem}
\usepackage{afterpage}
\usepackage{ulem}
\usepackage{tensor}
\usepackage{soul}
\usepackage{xcolor,lipsum,subcaption}
\usepackage{booktabs,makecell}

\definecolor{dred}{rgb}{0.8,0,0.1}
\definecolor{orange}{rgb}{1,0.5,0}

\renewcommand{\arraystretch}{1.5}
\DeclareUnicodeCharacter{2212}{\textendash}
\begin{document}
\preprint{APS/123-QED}

\title{
 Towards a possible solution to the Hubble tension with Horndeski gravity
}

\author{Yashi Tiwari}%
\email{yashitiwari@iisc.ac.in}
\affiliation{%
Joint Astronomy Programme, Department of Physics, Indian Institute of Science, C.V. Raman Road,
Bangalore 560012, India
}%

\author{Basundhara Ghosh}
 \email{basundharag@iisc.ac.in}
 \affiliation{%
 Department of Physics, Indian Institute of Science, C.V. Raman Road,
Bangalore 560012, India
}%

\author{Rajeev Kumar Jain}%
 \email{rkjain@iisc.ac.in}
\affiliation{%
Department of Physics, Indian Institute of Science, C.V. Raman Road,
Bangalore 560012, India
}%

\begin{abstract}
The Hubble tension refers to the discrepancy in the value of the Hubble constant $H_0$ inferred from the cosmic microwave background observations, assuming the concordance $\Lambda$CDM model of the Universe, and that from the distance ladder and other direct measurements. In order to alleviate this tension, we construct a plausible dark energy scenario, within the framework of Horndeski gravity which is one of the most general scalar-tensor theories yielding second-order equations. In our set-up, we include the self-interactions and nonminimal coupling of the dynamical dark energy scalar field which enable very interesting dynamics leading to a phantom behaviour at low redshifts along with negative dark energy densities at high redshifts. These two features together make this model a promising scenario to alleviate the Hubble tension for appropriate choices of the model parameters. Towards a consistent model building, we show that this set-up is also free from both the gradient and ghost instabilities. Finally, we confront the predictions of the model with low redshift observations from Pantheon, SH0ES, cosmic chronometers and BAO, to obtain best fit constraints on model parameters.
\end{abstract}

                    
\maketitle
\section{Introduction}
\label{sec:intro}

One of the most significant discoveries in modern cosmology is the observational evidence of the accelerated expansion of the present Universe, which has been independently inferred from the observations of distant type Ia supernovae \cite{SupernovaSearchTeam:1998fmf,SupernovaCosmologyProject:1998vns}. Such an accelerated expansion is thought to be caused by an unknown yet dominant component of the Universe called dark energy. Within the regime of general relativity (GR), the Standard Model of Cosmology, also known as the $\Lambda$-Cold Dark Matter ($\Lambda$CDM) model, can explain this phenomenon with the help of a cosmological constant $\Lambda$. Despite being the simplest model describing the Universe, the $\Lambda$CDM framework is plagued by fundamental issues like the cosmological constant problem and lack of physical understanding regarding the late time acceleration of the Universe, which might be arising due to some unknown dynamics at large scales. There are mainly two approaches that have been followed in order to address these issues -- either by considering GR to be the accepted theory of gravity and modifying the dark energy components \cite{Copeland:2006wr} or by modifying the gravitational theory itself in a way that leads to an accelerated expansion of the Universe, while reducing to GR in certain limits \cite{Clifton:2011jh}.

In the former case, one disregards the cosmological constant $\Lambda$ and instead introduces a scalar field that mimics the dynamical dark energy, thereby leading to the late time accelerated expansion.  On the other hand, a modification of the underlying gravitational theory is usually done by introducing some extra degrees of freedom (DoF) in addition to Einstein's gravity or GR. One of the ways to manifest this is via the scalar-tensor theories of gravity \cite{Fujii:2003pa} wherein the extra DoF are mixed with curvature, and on subjecting the action to a conformal transformation, they get non-minimally coupled with the matter. The idea is that one should be able to make use of such modified gravity theories to generate extra DoF that mimic the late time acceleration at large scales and hence resolve the problems that the $\Lambda$CDM model faces. However, with respect to the local measurements, these DoF should be suppressed so that Einstein's gravity, which agrees very well with observations, becomes the underlying theory of gravity. The required suppression of such extra DoF so as to be consistent with local measurements is usually achieved via various screening mechanisms such as the chameleon \cite{Khoury:2003rn} and the Vainshtein mechanisms \cite{Vainshtein:1972sx}.

Over the years, several discrepancies in the standard $\Lambda$CDM model have also emerged with the advancement of precision cosmology. These are related to the mismatch in the values of cosmological parameters obtained from direct observations as compared to those inferred from the concordance $\Lambda$CDM model. One of the most interesting and statistically significant discrepancies is in the value of Hubble constant $H_0$ inferred from the cosmic microwave background (CMB) data \cite{Planck:2018vyg} which assumes a $\Lambda$CDM model and those obtained from distance ladder measurements conducted by the SH0ES collaboration \cite{Riess:2021jrx}. Although one can attribute this mismatch to the systematic errors in direct measurement of $H_0$, multiple observations from alternative methods have also indicated a tension with $H_0$ inferred from $\Lambda$CDM \cite{Freedman:2021ahq,Anand:2021sum,deJaeger:2022lit,Blakeslee:2021rqi,Pesce:2020xfe,FernandezArenas:2017dux,Shajib:2023uig}. At present, the observations from SH0ES give a high value of  $H_0=(73.3\pm 1.04)$ km/s/Mpc, with a $5\sigma$ tension with that inferred from Planck. The Hubble tension thus raises a direct question about our understanding of the expansion of the Universe, thereby emphasizing the need to look for possible modifications in the standard $\Lambda$CDM picture.
Apart from the $H_0$ tension,  the $\sigma_8$ (the cosmological parameter related to the clustering of matter in an $8h^{-1}$Mpc radius sphere) tension refers to a discrepancy between the estimation from the CMB and the measurement from surveys like BOSS \cite{BOSS:2016wmc, Philcox:2021kcw}.  Some other discrepancies are associated with the Ly-$\alpha$ baryon acoustic oscillations (BAO), which provide an independent measurement of the expansion history at higher redshifts. The Ly-$\alpha$ measurement from eBOSS at $z \sim 2.33$ prefers a lower value of $H(z)$ than $\Lambda$CDM model \cite{duMasdesBourboux:2020pck}, although with a mild statistical significance of 1.5$\sigma$.

Recently, a large number of attempts have been made to alleviate the Hubble tension using different methods. A comprehensive summary of these approaches can be found in \cite{Verde:2019ivm, Freedman:2021ahq, Knox:2019rjx, DiValentino:2021izs,Perivolaropoulos:2021jda,Zumalacarregui:2020cjh,Abdalla:2022yfr,Shah:2021onj,Schoneberg:2021qvd}, most of which can be categorised either as an early-time or a late-time modification. For a particular cosmological model, one can obtain the value of $H_0$ via a measurement of the angular scale of sound horizon $\theta_s=r_s(z_{\text{rec}})/d_A(z_{\text{rec}})$, where $z_{\text{rec}}$ is the redshift at recombination, $r_s(z_{\text{rec}})$ is the sound horizon, and $d_A(z_{\text{rec}})$ is the corresponding angular diameter distance. Since $\theta_s$ is constrained to high precision by CMB data \cite{Planck:2018vyg}, various solutions to the $H_0$ tension must necessarily respect this constraint. The early-time solutions usually employ a reduction in $r_s(z_{\text{rec}})$, which would require increasing $d_A(z_{\text{rec}})$ in order to keep $\theta_s$ fixed, automatically implying an increase in the value of $H_0$. Some of these approaches include combining the BAO and the Big Bang Nucleosynthesis (BBN) data \cite{Schoneberg:2019wmt, Schoneberg:2022ggi}, imposing constraints on early dark energy \cite{Karwal:2016vyq, Agrawal:2019lmo, Smith:2019ihp, Lin:2019qug, Poulin:2018cxd, Niedermann:2019olb, Niedermann:2020dwg, Herold:2022iib}, hot new early dark energy \cite{Niedermann:2021ijp, Niedermann:2021vgd}, cascading dark energy \cite{Rezazadeh:2022lsf}, early dark energy with thermal friction \cite{Berghaus:2019cls, Berghaus:2022cwf}, and using massive neutrinos \cite{Poulin:2018zxs, Kreisch:2019yzn, Sakstein:2019fmf}. However, these approaches encounter a myriad of problems, and it is understood that early-time solutions are not sufficiently capable of alleviating the $H_0$ tension \cite{Vagnozzi:2023nrq}. 

On the other hand, late-time solutions incorporate mechanisms to alter the rate of expansion at redshifts close to the present, which inherently implies $z \ll z_{\text{rec}}$, thus resulting in a higher value of $H_0$ without disturbing either $r_s(z_{\text{rec}})$ or $d_A(z_{\text{rec}})$, and hence $\theta_s$. The usual prototype to bring about this is to work with a phantom dark energy model, where the equation of state behaves as $w<-1$ instead of the $\Lambda$CDM model (with $w=-1$). To this end, a few of the numerous attempts include employing ``quintessence" potentials to solve the $H_0$ and $\sigma_8$ tensions simultaneously \cite{Adil:2022hkj}, considering a quintessence field that tracks the equation of state during recombination and then boosts the late-time expansion \cite{DiValentino:2019exe}, tuning the late-time acceleration of the Universe to cause a monotonous increase in the expansion rate \cite{Adil:2021zxp}, employing a phenomenological two-parameter family of dark energy models \cite{Lee:2022cyh},  interacting dark energy models \cite{DiValentino:2017iww,Gariazzo:2021qtg,Yao:2023jau}, using new gravitational scalar-tensor theories that are free of ghosts by definition \cite{Banerjee:2022ynv}, and many more \cite{Alexander:2019rsc, Krishnan:2020obg, Adi:2020qqf, Braglia:2020iik}.

In this paper, we introduce a plausible dark energy scenario to provide a possible resolution of the Hubble tension, which is motivated within the framework of Horndeski gravity \cite{Horndeski:1974wa} -- one of the most general scalar-tensor theories involving second-order field equations. 
Horndeski theory provides a rich phenomenology to modify the underlying gravitational theory by introducing an additional degree of freedom, a scalar field, leading to interesting implications in inflationary \cite{Chien:2021zle,Sebastiani:2017mkv,Hikmawan:2019idy, Tiwari:2022zzz} well as dark energy physics \cite{Kennedy:2017sof,Bayarsaikhan:2020jww,Kase:2018aps}. Recently Horndeski theories have been exploited to address Hubble tension as well \cite{Petronikolou:2021shp,Banerjee:2022ynv,Petronikolou:2023cwu}. We consider a special case of the Horndeski setup with a dynamical dark energy scalar field $\phi$ involving self-interaction terms and nonminimal coupling. This leads to interesting features in the dynamics of the late universe and paves the way for a resolution of the Hubble tension. We emphasize two important characteristics of the model, the first being phantom behavior (implying that the dark energy equation of state $w_\phi<-1$) at present while the other is negative dark energy density ($\rho_\phi<0$) at high redshifts. The phantom dark energy models are well motivated in literature to provide a faster expansion phase leading to large $H_0$, thereby resolving the Hubble tension \cite{Li:2019yem,Alestas:2020mvb,Yang:2018qmz,DiValentino:2016hlg,Vagnozzi:2019ezj,DiValentino:2019dzu}, but at the same time being disfavoured by BAO, $\sigma_8$ (or $S_8$) measurements, unless accompanied by a phantom crossing behaviour at some epoch \cite{Heisenberg:2022lob,Abdalla:2022yfr,DiValentino:2020naf}. On the other hand, a negative dark energy density at high redshifts has been preferred by various data and observational reconstructions \cite{Aubourg:2014yra,Poulin:2018zxs,Wang:2018fng,Bonilla:2020wbn,Malekjani:2023dky,Mehrabi:2022ywh,Dutta:2018vmq,Akarsu:2022lhx}, and has been found to be helpful in resolving many anomalies and cosmological tensions like BAO Ly-$\alpha$ measurements, $H_0$ and $S_8$ tensions \cite{Sahni:2014ooa,Akarsu:2019ygx,Visinelli:2019qqu,Akarsu:2019hmw,Akarsu:2022typ,Sen:2021wld,Adil:2023exv}.

In this work, we show that with the combined features mentioned above, our model promises to effectively address the $H_0$ tension. We also comment upon the predictions of the model towards simultaneous alleviation of $H_0$ and $\sigma_8$ tensions based on the discussion by Heisenberg et al. \cite{Heisenberg:2022lob}. The dark energy scenario proposed in this work is a late-time modification in the dynamics of the universe, thereby not affecting the early universe physics before recombination. Towards a consistent, well-behaved theory, we show that the model is free of gradient and ghost instabilities, which may otherwise arise in more general scalar-tensor theories. Finally, we compare the model predictions against low redshift data from Pantheon, SH0ES, BAO, and cosmic chronometers (CC) \cite{Pan-STARRS1:2017jku,Riess:2021jrx,Yu:2017iju,Ata:2017dya} and employ a Markov Chain  Monte Carlo (MCMC) analysis to obtain best fit constraints on model parameters.

Our paper is organised as follows: in Section \ref{sec:horndeski_intro}, we briefly introduce the Horndeski theory and its dynamics and state the conditions for the avoidance of any gradient and ghost instabilities. In Section \ref{sec:model}, we present our model as a special case of Horndeski Lagrangian with a particular choice of interaction terms. In this section, we further discuss in detail the dynamics and features of the model and confront the model predictions with available data. Finally, in Section \ref{sec:conclusion}, we summarize our results and conclude with some discussions and future directions.

We work with natural units such that $\hbar=c=1$, and set the reduced Planck mass $M_\text{Pl}=(8\pi G)^{-1/2}=1$. Our metric signature is ~$(-,+,+,+)$.
\section{Horndeski theory and its dynamics}
\label{sec:horndeski_intro}


We work in the framework of the Horndeski theory which is a generalised scalar-tensor theory in four dimensions. The Lagrangian is constructed out of the metric tensor and a scalar field and leads to second order equations of motion, thereby being free of Ostrogradsky instability. The complete Lagrangian for the Horndeski theory (or equivalently, for the generalized Galileons) is given by 
\begin{equation}
\mathcal{L}= \sum_{i=2}^5 \mathcal{L}_{i}\,,
\label{e: horndeski_lag}
\end{equation}

where
\begin{eqnarray}
     && \mathcal{L}_2\ =\  G_2(\phi, X),\nonumber\\
     && \mathcal{L}_3\ =\  -G_3(\phi, X)\Box{\phi},\nonumber\\
     && \mathcal{L}_4\ =\  G_4(\phi, X)R + G_{4,X}(\phi, X)\Big[(\Box{\phi})^2 - (\nabla_{\mu}\nabla_{\nu}\phi)^2\Big],\nonumber\\
     && \mathcal{L}_5\ =\  G_5(\phi,X)G_{\mu\nu}\nabla^{\mu}\nabla^{\nu}\phi - \frac{1}{6}  G_{5,X}(\phi, X)\Big[(\Box{\phi})^3-3\Box{\phi}(\nabla_{\mu}\nabla_{\nu}\phi)^2+2(\nabla_{\mu}\nabla_{\nu}\phi)^3\Big], \nonumber
\end{eqnarray}
where $R$ is the Ricci scalar, $G_i$ are four independent arbitrary functions of $\phi$ and $X$, and $X=-\partial_{\mu}\phi\partial^{\mu}\phi$/2, $G_{i,Y} =\partial{G}_i/\partial{Y}$ with $Y=\{\phi,X\}$.
Thus, in the Horndeski gravity, the complete action can be given as,
\begin{equation}
\mathcal{S} = \int \mathrm{d}^4x \sqrt{-g}\, \left(\sum_{i=2}^5 \mathcal{L}_{i}+\mathcal{L}_M\right)\,,
\label{e:horndeski_action}
\end{equation}
where,
$\mathcal{L}_M$ accounts for the matter (and the radiation) component. Such a theory can well explain the dark energy phase of Universe modelled by a scalar field $\phi$ coupled to gravity (minimal or non-minimal coupling) depending on the choice of $G_i(\phi,X)$. In a flat FRW background with $ds^2=-dt^2+a(t)^2 d\Bar{x}^2$, two Friedmann equations are given by \cite{DeFelice:2011bh},
\begin{eqnarray}
    2X G_{2,X}-G_2+6X\dot\phi H G_{3,X}-2 X G_{3,\phi}-6H^2 G_4+24H^2 X(G_{4,X}+X G_{4,XX}-12H X \dot\phi G_{4,\phi X} \nonumber\\
    -6H \dot\phi G_{4,\phi}+2H^3 X \dot\phi(5G_{5,X}+2X G_{5,XX})-6H^2 X(3G_{5,\phi}+2X G_{5,\phi X})=-\rho_M
\label{e:friedmann_first}
\end{eqnarray}
    
\begin{eqnarray}
    & G_{2}-2X(G_{3,\phi}+\ddot\phi G_{3,X})+2(3H^2+2\dot H)G_4-12H^2 X G_{4,X}-4H \dot X G_{4,X}-8\dot H X G_{4,X}\nonumber\\& -8H X\dot X G_{4,XX} +2(\ddot \phi+2H \dot \phi)G_{4,\phi}+4X G_{4,\phi\phi}+4X(\ddot \phi-2H\dot \phi)G_{4,\phi X}-2X(2H^3 \dot \phi+ 2H \dot H \dot \phi +3H^2 \ddot\phi)G_{5,X}\nonumber\\ & -4H^2 X^2 \ddot\phi G_{5,XX}+ 4H X(\dot X-H X)G_{5,\phi X}+2[2(\dot H X+H \dot X)+3H^2 X]G_{5,\phi}+4H X \dot\phi G_{5,\phi \phi}=-p_M
\label{e:friedmann_second}
\end{eqnarray}

In fact, the Eqs. \ref{e:friedmann_first} and \ref{e:friedmann_second} can be written in a simplified manner as \cite{Matsumoto:2017qil},

\begin{equation}
    3H^2= \kappa^2(\rho_{\phi}+\rho_M)
    \label{e:friedman-first}
\end{equation}
\begin{equation}
    -3H^2-2 \dot{H} = \kappa^2(p_\phi+p_M)
    \label{e:friedman-second}
\end{equation}

where $\rho_\phi$ and $p_\phi$ are the energy density and pressure of the scalar field given as,
\begin{eqnarray}
    \rho_\phi=2X G_{2,X}-G_2+6X\dot\phi H G_{3,X}-2 X G_{3,\phi}-6H^2 G_4+24H^2 X(G_{4,X}+X G_{4,XX}-12H X \dot\phi G_{4,\phi X} \nonumber\\
    -6H \dot\phi G_{4,\phi}+2H^3 X \dot\phi(5G_{5,X}+2X G_{5,XX})-6H^2 X(3G_{5,\phi}+2X G_{5,\phi X})+\frac{3 H^2}{\kappa^2}
    \label{e:rho-phi}
\end{eqnarray}

\begin{eqnarray}
     p_\phi=G_{2}-2X(G_{3,\phi}+\ddot\phi G_{3,X})+2(3H^2+2\dot H)G_4-12H^2 X G_{4,X}-4H \dot X G_{4,X} \nonumber\\  -8\dot H X G_{4,X}-8H X\dot X G_{4,XX} +2(\ddot \phi+2H \dot \phi)G_{4,\phi}+4X G_{4,\phi\phi}+4X(\ddot \phi-2H\dot \phi)G_{4,\phi X}\nonumber\\ -2X(2H^3 \dot \phi+ 2H \dot H \dot \phi +3H^2 \ddot\phi)G_{5,X} -4H^2 X^2 \ddot\phi G_{5,XX}+ 4H X(\dot X-H X)G_{5,\phi X} \nonumber\\ +2[2(\dot H X+H \dot X)+3H^2 X]G_{5,\phi}+4H X \dot\phi G_{5,\phi \phi} -\frac{1}{\kappa^2} (3H^2+ 2 \dot H)
\label{e:pressure-phi}
\end{eqnarray}

In the above equations, $\kappa^2=1/M_\text{Pl}^2$, and from now on, we set  $\kappa^2=1$.  Finally, the equation for the evolution of the scalar field is given by \cite{DeFelice:2011bh},
    
\begin{equation}
    \frac{1}{a^3}\frac{d}{dt}(a^3
    \mathcal{J})=\mathcal{P}_\phi
\label{e:phi_evolution}
\end{equation}
where,
\begin{eqnarray}
    \mathcal{J}=\dot\phi G_{2,X}+6H X G_{3,X}-2 \dot\phi G_{3,\phi}+6H^2 \dot\phi(G_{4,X}+2X G_{4,XX})-12H X G_{4,\phi X} \nonumber\\
    +2H^3 X(3G_{5,X}+2X G_{5,XX})-6H^2 \dot\phi(G_{5,\phi}+X G_{5,\phi X}),
\label{e:J}
\end{eqnarray}

\begin{eqnarray}
    \mathcal{P}_\phi=G_{2,\phi}-2X(G_{3,\phi\phi}+\ddot\phi G_{3,\phi X})+6(2H^2+\dot H)G_{4,\phi}+6H(\dot X+2H X)G_{4,\phi X}\nonumber\\
    -6H^2 X G_{5,\phi \phi}+2H^3 X \dot \phi G_{5,\phi X}.
\label{e:P_phi}
\end{eqnarray}
In the above equations $\rho_M$ and $p_M$ are the energy density and pressure of the matter field (and radiation), which satisfy the continuity equation,
\begin{equation}
    \dot \rho_M+3 H \rho_M (1+w_M)=0,
\label{e:continuity}
\end{equation}
where $w_M=p_M/\rho_M$ is the equation of state for the fluid. The complete background evolution of the Universe can be obtained by solving the above set of equations. However, due to the non-trivial and non-canonical structure of the Horndeski Lagrangian, sometimes these models can have instability issues in the evolution of the perturbations, rendering the background evolution inappropriate. It is therefore necessary to keep a check on the parameters of the theory, even if one is not concerned with the evolution of perturbations for a given dark energy model. The Laplacian or gradient instability is related to the propagation speed of the scalar and tensor perturbations which arise in the regime where the square of sound speed of perturbations becomes negative. This leads to an unstable growth of the perturbation modes on small scales. Another crucial one is the ghost instability which arises when the sign of the kinetic term of the perturbations goes negative. We shall work in the parameter space wherein we avoid all these instabilities so as to have a well-behaved theory. Following the standard technique of linear cosmological perturbation theory, one can obtain the second-order action for the scalar and tensor perturbations in Horndeski theory as \cite{DeFelice:2011bh},
\begin{equation}
    \mathcal{S}_2=\int \mathrm{d}t \mathrm{d}^3 x a^3 \left[Q_S\left(\mathcal{\dot R}^2-\frac{c_S^2}{a^2}(\partial_i \mathcal{R})^2\right)+Q_T\left(\dot h_{ij}^2-\frac{c_T^2}{a^2}(\partial_k h_{ij})^2\right)\right],
\label{e:action_scalartensor}
\end{equation}
where $\mathcal{R}$ is the scalar curvature perturbation while $h_{ij}$ are the tensor modes or gravitational waves (GWs), and $Q_S$ and $Q_T$ are given by
\begin{equation}
    Q_S \equiv \frac{w_1\left(4 w_1 w_3+9 w_2^2\right)}{3 w_2^2}
\end{equation}
\begin{equation}
    Q_T \equiv \frac{w_1}{4}
\end{equation}
Similarly, $c_S$ and $c_T$ are the propagation speeds of the scalar and tensor modes, respectively, and are given as,
\begin{eqnarray}
   c_S^2 \equiv \frac{3\left(2 w_1^2 w_2 H-w_2^2 w_4+4 w_1 w_2 \dot{w}_1-2 w_1^2 \dot{w}_2\right)-6 w_1^2\left[\left(1+w_A\right) \rho_A+\left(1+w_B\right) \rho_B\right]}{w_1\left(4 w_1 w_3+9 w_2^2\right)}
\label{e:cssqr}
\end{eqnarray}
and 
\begin{equation}
    c_T^2 \equiv \frac{w_4}{w_1}
    \label{e:cTsqr}
\end{equation}
where,
\begin{equation}
w_1 \equiv 2\left(G_4-2 X G_{4, X}\right)-2 X\left(G_{5, X} \dot{\phi} H-G_{5, \phi}\right),
\label{e:w1}
\end{equation}
\begin{eqnarray}
w_2 &\equiv & -2 G_{3, X} X \dot{\phi}+4 G_4 H-16 X^2 G_{4, X X} H+4\left(\dot{\phi} G_{4, \phi X}-4 H G_{4, X}\right) X+2 G_{4, \phi} \dot{\phi}\nonumber\\
& +& 8 X^2 H G_{5, \phi X}+2 H X\left(6 G_{5, \phi}-5 G_{5, X} \dot{\phi} H\right)-4 G_{5, X X} \dot{\phi} X^2 H^2,
\label{e:w2}
\end{eqnarray}
\begin{eqnarray}
w_3 &\equiv & 3 X\left(K_{, X}+2 X K_{, X X}\right)+6 X\left(3 X \dot{\phi} H G_{3, X X}-G_{3, \phi X} X-G_{3, \phi}+6 H \dot{\phi} G_{3, X}\right) \nonumber\\
& +&18 H\left(4 H X^3 G_{4, X X X}-H G_4-5 X \dot{\phi} G_{4, \phi X}-G_{4, \phi} \dot{\phi}+7 H G_{4, X} X+16 H X^2 G_{4, X X}-2 X^2 \dot{\phi} G_{4, \phi X X}\right) \nonumber\\
& +& 6 H^2 X\left(2 H \dot{\phi} G_{5, X X X} X^2-6 X^2 G_{5, \phi X X}+13 X H \dot{\phi} G_{5, X X}-27 G_{5, \phi X} X+15 H \dot{\phi} G_{5, X}-18 G_{5, \phi}\right), \nonumber\\
\label{e:w3}
\end{eqnarray}
\begin{equation}
   w_4 \equiv 2 G_4-2 X G_{5, \phi}-2 X G_{5, X} \ddot{\phi} .
   \label{e:w4}
\end{equation}
For consistent dynamics, free of any gradient and ghost instabilities for both scalar and tensor modes, one must satisfy

\begin{eqnarray}
    c_S^2>0, \hspace{.5cm} Q_s>0, \hspace{.5cm} c_T^2>0, \hspace{.5cm} Q_T>0
    \label{e: instability_check}
\end{eqnarray}

Another non-trivial characteristic of Horndeski theories is that the speed of GWs can evolve with time for models containing $G_4(\phi,X)$ and $G_5(\phi,X)$, as can be seen from Eq. \ref{e:cTsqr} and in fact using Eqs. \ref{e:w1} and \ref{e:w4}, can be stated as 
\begin{equation}
   c_{T}^2= \frac{G_4-X G_{5,\phi}-X G_{5,X}\Ddot{\phi}}{G_4-2X G_{4,X}-X(G_{5,X} \dot\phi H-G_{5,\phi})}
   \label{e:cTsqr}
\end{equation}

The recent observations of GWs from the LIGO-VIRGO collaboration and their electromagnetic counterparts put a very stringent bound on speed of GWs, such that
\cite{Gong:2017kim, LIGOScientific:2017zic}
\begin{align}
-3\times 10^{-15}<c_T-1<7\times10^{-16} \,\label{e:cTbound}
\end{align}
which implies that a consistent dark energy model should not violate Eq. \ref{e:cTbound} in any regime. 


\section{Our scenario with self-interactions and non-minimal coupling}
\label{sec:model}
In this section, we present a model constructed within the framework of the Horndeski Lagrangian, comprising of non-zero $G_2$, $G_3$, and $G_4$ terms of Eqs. \ref{e: horndeski_lag} such that,
\begin{equation}
    G_2=X-V(\phi),\quad G_3=c_1\phi+c_2 X,\quad G_4=\frac{1}{2}+c_3\phi,\quad G_5=0\,.
\label{e:setup}
\end{equation}
Here we choose, $G_4$ as a function of the dark energy scalar field $\phi$ i.e. $G_4(\phi,X)=G_4(\phi)$ and $G_5(\phi,X)=0$. This choice ensures that the speed of gravitational waves is always luminal, i.e. $c_{T}=1$, as can be verified easily by Eq. \ref{e:cTsqr}. Our model as specified by Eq. \ref{e:setup} can also be seen as an extension to the quintessence dark energy scenario \cite{Tsujikawa:2013fta}, where
$\phi$ is the scalar field that drives the dark energy phase, $X=\dot\phi^2/2$ is the kinetic term, and $V=V(\phi)$ is the potential of the field. Clearly, $G_3=0$, $G_4=1/2$ corresponds to the quintessence case, while $\Lambda$CDM scenario can be obtained by choosing $G_2=-2\Lambda$,  $G_3=0$ and $G_4=1/2$.
\par

For the model given by Eq. \ref{e:setup}
the energy density and pressure of the dynamical scalar field can be written using Eqs. \ref{e:rho-phi} and \ref{e:pressure-phi} as,

\begin{equation}
    \rho_\phi= \frac{1}{2}\dot\phi^2+V(\phi)-6 c_3 \phi H^2 - 6 c_3 H \dot\phi - c_1 \dot\phi^2 + 3 c_2 H \dot\phi^3  
    \label{e:rho-ourmodel}
\end{equation}

\begin{equation}
    p_\phi= \frac{1}{2}\dot\phi^2-V(\phi)+6 c_3 \phi H^2+ 4 c_3 \phi \dot H+2 c_3 \Ddot{\phi}+4 c_3 H \dot\phi-c_1 \dot\phi^2 -c_2 \Ddot{\phi} \dot\phi^2
    \label{e:p-ourmodel}
\end{equation}

Therefore, using the above expressions for $\rho_\phi$ and $p_\phi$ in Eqs. \ref{e:friedmann_first} and \ref{e:friedmann_second} one can obtain the two Friedman equations. Further, the evolution of the dark energy scalar field for our setup of Eq. \ref{e:setup} can be written using Eq. \ref{e:phi_evolution} as,

\begin{equation}
    \Ddot{\phi}+3 H \dot\phi+ V'(\phi)-12 c_3 H^2-6 c_3 \dot H-2 c_1 \Ddot{\phi}-6 c_1 H \dot\phi+6 c_2 H \Ddot{\phi} \dot\phi+9 c_2 H^2 \dot\phi^2+3 c_2 \dot H \dot\phi^2=0
    \label{e:phi-ourmodel}
\end{equation}

Finally, Eqs. \ref{e:friedman-first},\ref{e:friedman-second} and \ref{e:phi-ourmodel} form the set of three equations that provide the evolution of the cosmological background in the presence of the dynamical dark energy scalar field with energy density and pressure given by Eqs. \ref{e:rho-ourmodel} and \ref{e:p-ourmodel}, respectively. These equations are solved numerically to obtain the evolution of relevant background quantities such as the Hubble parameter $H(z)$, dark energy equation of state $w_\phi(z)$, etc. Throughout the analysis, we fix the potential of the dark energy scalar field to a linear one, i.e., $V(\phi)=V_0 \phi$.
The initial conditions on $H$, $\phi$, and $\dot\phi$ are set at the time of recombination, i.e. at $z_{\rm rec}=1100$. We set $H(z_{\rm rec})=H_{\rm \Lambda CDM}(z_{\rm rec})$ in $H_0$ units, while we choose $\phi(z_{\rm rec})=2$ and $\dot\phi(z_{\rm rec})=10^{-3}$ in Planck units. Also, the parameter $V_0$ (of dimension \small{$[M_\text{Pl}]^3$}) of the dark energy potential is kept fixed throughout the analysis, with $V_0=1.1$ in $H_0$ units.

The three crucial model parameters in our analysis are $c_1$, $c_2$, and $c_3$, which control the strength of couplings as can be seen from Eq. \ref{e:setup}. Specifically, $c_1$ and $c_2$ contained inside $G_3$ control the self-interaction terms of the scalar field and its derivative, while $c_3$ in $G_4$ controls the strength of nonminimal coupling to gravity. These two interactions arising from $G_3$ and $G_4$, respectively, lead to distinct effects in the dynamics of the expansion of the universe. Specifically, the $G_3$ term in Eq. \ref{e:setup}, with an appropriate choice of $c_1$ and $c_2$ can give a phantom behavior ($w_{\phi}<-1$) at low redshifts leading to a faster expansion compared to $\Lambda$CDM, while the nonminimal coupling to gravity due to the choice of $G_4$ in Eq. \ref{e:setup}, leads to negative dark energy density at high redshifts (around $z \geq 3$). This can be qualitatively seen from Eq. \ref{e:rho-ourmodel}, where the appropriate choice of $c_3 (>0)$ can lead to $\rho_{\phi}<0$ in regimes where $V(\phi)$ is subdominant, given that $\dot\phi$ is really small. In Sec. \ref{sec:appendix}, we shall discuss these effects in detail. As mentioned in Sec. \ref{sec:intro}, negative dark energy at high redshifts has been found to be effective in resolving anomalies and cosmological tensions. In the next section, we will demonstrate these effects for different choices of $c_1,c_2$, and $c_3$. Also, the dimensions of the three model parameters, $c_1$, $c_2$, $c_3$ in terms of Planck mass are \small{$[M_\text{Pl}]^0$}, \small{$[M_\text{Pl}]^{-3}$} and \small{$[M_\text{Pl}]^1$} respectively. The case with $c_1=c_2=c_3=0$ gives the usual quintessence scenario, with a background similar to the $\Lambda$CDM.

Another interesting feature of the Horndeski theory involving a non-trivial $G_4$ term is that it can lead to an evolution of the Planck mass on cosmological scales. This leads to a modification in the luminosity distance compared to the standard GR due to additional friction arising from the running Planck mass \cite{Linder:2021pek}, referred to as the GW distance. Future GW distance observations from the Einstein Telescope will constrain the deviations from GR, arising in modified gravity theories like Horndeski gravity \cite{Matos:2022uew}. Furthermore, the presence of Galileon-type interaction in our model, i.e. $G_3(X)$, leads to the Vainshtein mechanism, thereby reconciling with GR inside the Vainshtein radius \cite{Ferrari:2023qnh}.

\subsection{Alleviating the Hubble tension -- our model results}

\begin{figure}[t]
    \centering
    \includegraphics[width=0.48\textwidth]{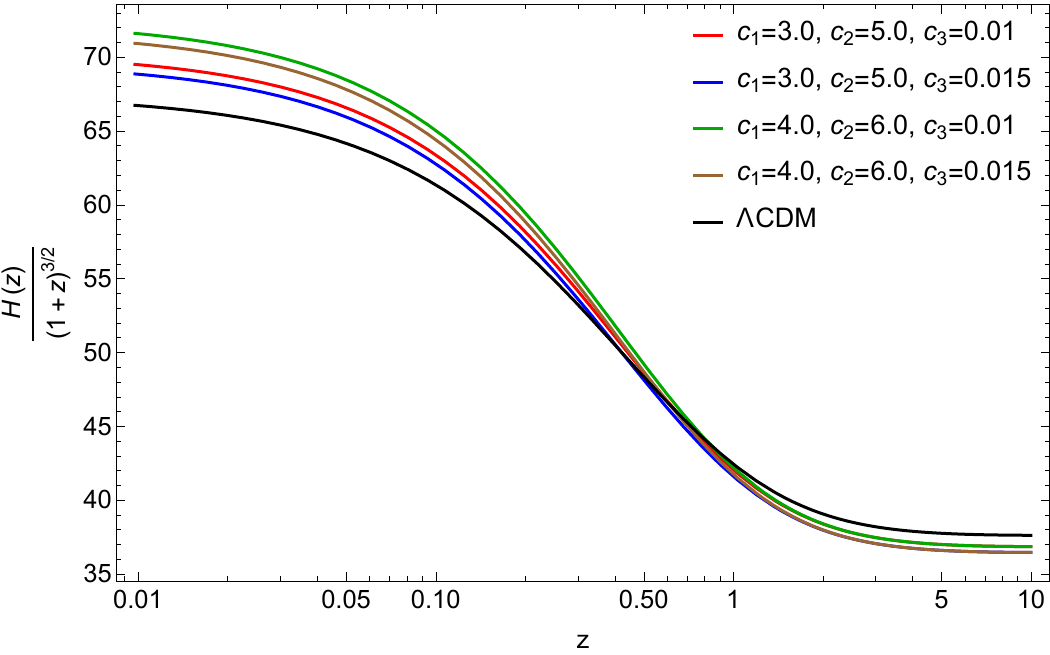}
    \includegraphics[width=0.48\textwidth]{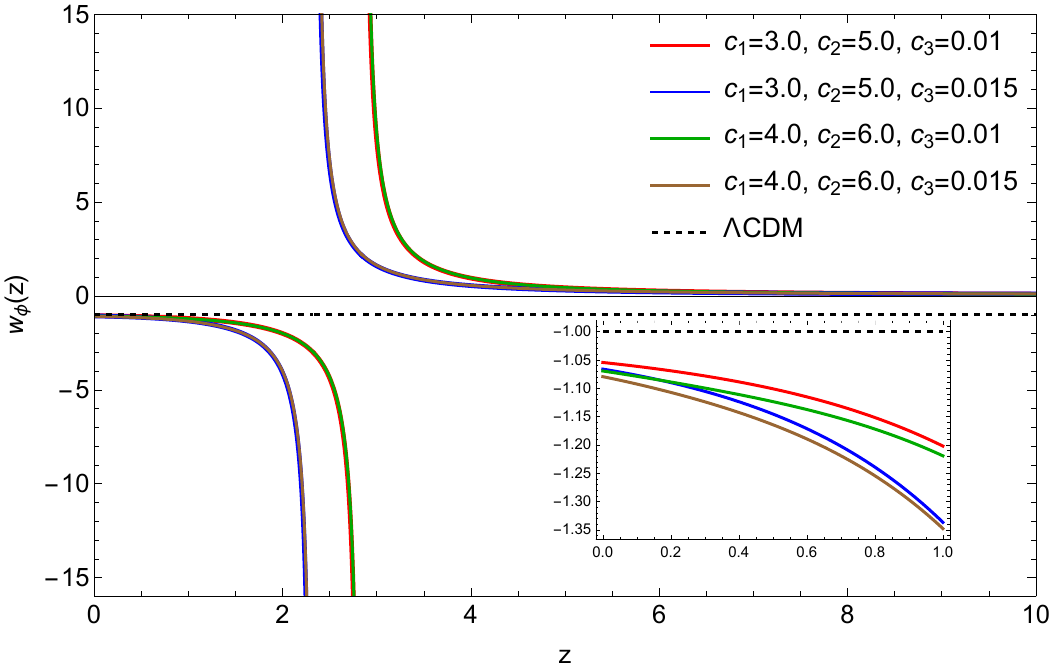}
    \includegraphics[width=0.48\textwidth]{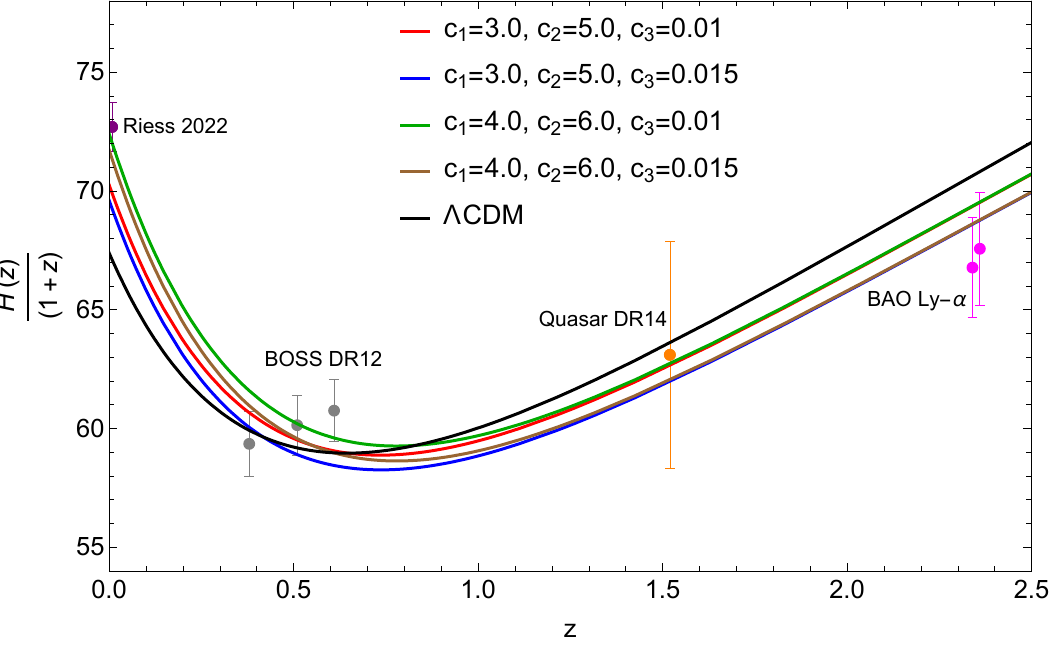}
     \includegraphics[width=0.48\textwidth]{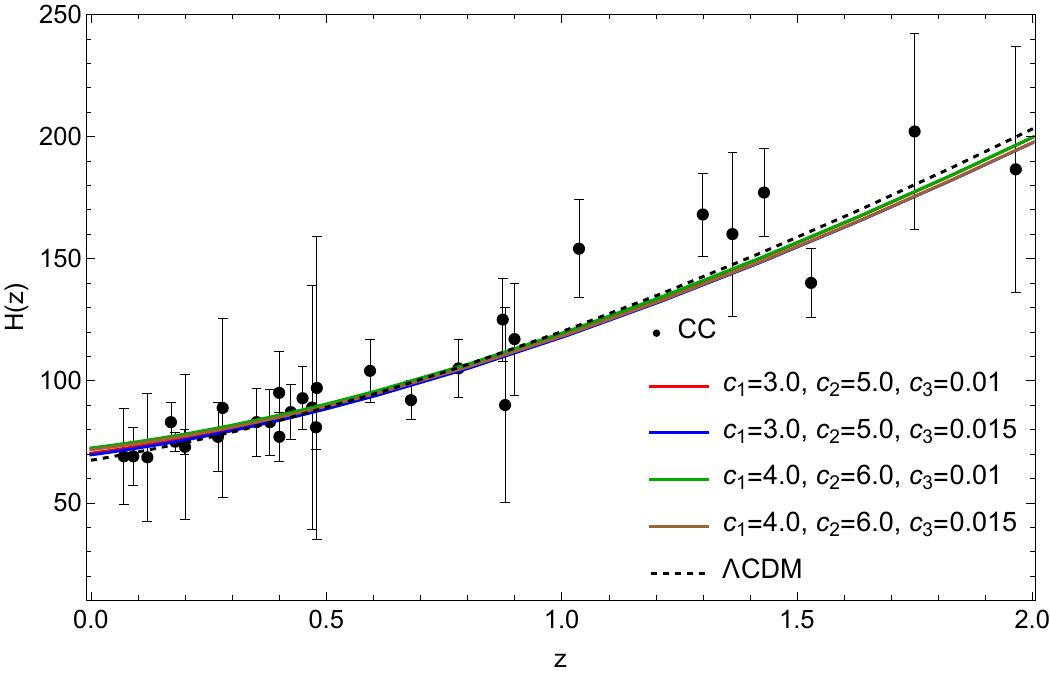}
    \caption{The evolution of the normalized Hubble parameter (top left) and equation of state of dark energy scalar field (top right) for different combinations of $c_1$, $c_2$, and $c_3$. In the bottom panel, the model is confronted with the measurement of $H(z)$ from SN \cite{Riess:2021jrx}, BAO, and CC data \cite{Yu:2017iju, Ata:2017dya}.}
    \label{fig:all_on}
\end{figure}

In this section, we present the results of the numerical analysis of the cosmological background for our model given by Eq. \ref{e:setup}. As mentioned earlier, we fix the initial conditions on $\phi$, $\dot\phi$, and $H$ at recombination $z_{\rm rec}$, along with a fixed value of $V_0$, to solve the set of background equations numerically. The top left panel of Fig. \ref{fig:all_on} shows the evolution of the normalized Hubble parameter $H(z)$ for different values of $c_1, c_2$ and $c_3$ while the top right panel shows the corresponding evolution of the dark energy equation of state $w_\phi(z)$. The results of the standard $\Lambda$CDM model are also displayed for comparison. In order to visualize the salient features of the model, we also confront the corresponding evolution of the Hubble parameter with observational data from BAO (bottom left) and cosmic chronometers (bottom right), as shown in Fig. \ref{fig:all_on}. The model can favour larger values of the Hubble parameter at present in comparison to the $\Lambda$CDM case for specific choices of model parameters. This can be seen in the top left panel of Fig. \ref{fig:all_on} that the model gives a comparatively faster expansion at low redshifts than $\Lambda$CDM, thereby large $H_0$. This can clearly be attributed to the phantom behavior of the dark energy scalar ($w_{\phi}<-1$) field around the present epoch in the top right panel of Fig. \ref{fig:all_on}.

However, at higher redshifts, the $H(z)$ for the model is less than that of $\Lambda$CDM, which is due to the negative energy density of the scalar field around those redshifts. This feature of $\rho_{\phi}<0$ at high redshifts is traced in the evolution of $w_{\phi}$ in the right panel of Fig. \ref{fig:all_on}, which crosses a singularity around the epoch ($z \sim 2-5$) where $\rho_{\phi}$ switches sign. Around this point, the dark energy density transits to positive values, giving appropriate evolution at present. This type of singularity in $w_\phi(z)$, arising due to negative dark energy in some regime, has been studied in literature to address anomalies in Ly-$\alpha$ BAO measurements of $H(z)$, as well as supported by the observational reconstructions \cite{Sahni:2014ooa,Wang:2018fng,Akarsu:2019ygx}.  Recently it has been pointed out (\cite{Malekjani:2023dky,Akarsu:2022lhx} and references therein) that the negative energy density of dark energy scalar field $\rho_\phi$ at high redshifts, wherein the effects of dark energy are really subdominant or insignificant, can lead to resolution of cosmological tensions like $H_0$, $S_8$ etc.  Indeed, the total energy density of the universe is always positive, even when $\rho_\phi$ is negative at higher redshifts. Since the contribution of $\rho_\phi$ to the total energy density remains really insignificant at earlier epochs, the singularity in the equation of state does not affect the evolution of relevant background entities like $H, \dot H$, etc. Thereby the two crucial characteristics of the dark energy component in our model i.e. the phantom behavior at low redshifts and negative energy density at higher redshifts, collectively are capable of explaining the observed measurement of Hubble parameter from independent observations like Supernovae, BAO and CC as demonstrated in the lower panel of Fig. \ref{fig:all_on}. To illustrate further, these two characteristics of the model lead to a sign switching in $\delta H(z)$ ($ =H(z)-H_{\rm \Lambda CDM}(z)$), i.e. $\delta H(z)<0$ at high redshifts while $\delta H(z)>0$ around the present, which is seen as a necessary condition to be satisfied by any late dark energy model trying to address the Hubble tension \cite{Heisenberg:2022lob, Adil:2023exv}. These conditions are obtained to ensure that any late-time modifications do not disturb the precisely measured angular size of the sound horizon around recombination by Planck. In Sec. \ref{sec:appendix}, we shall further illustrate the distinct effects arising from the individual interaction terms, i.e. from $G_3$ and $G_4$ of Eq. \ref{e:setup}, one at a time.

\begin{figure}
    \centering
    \includegraphics[width=0.48\textwidth]{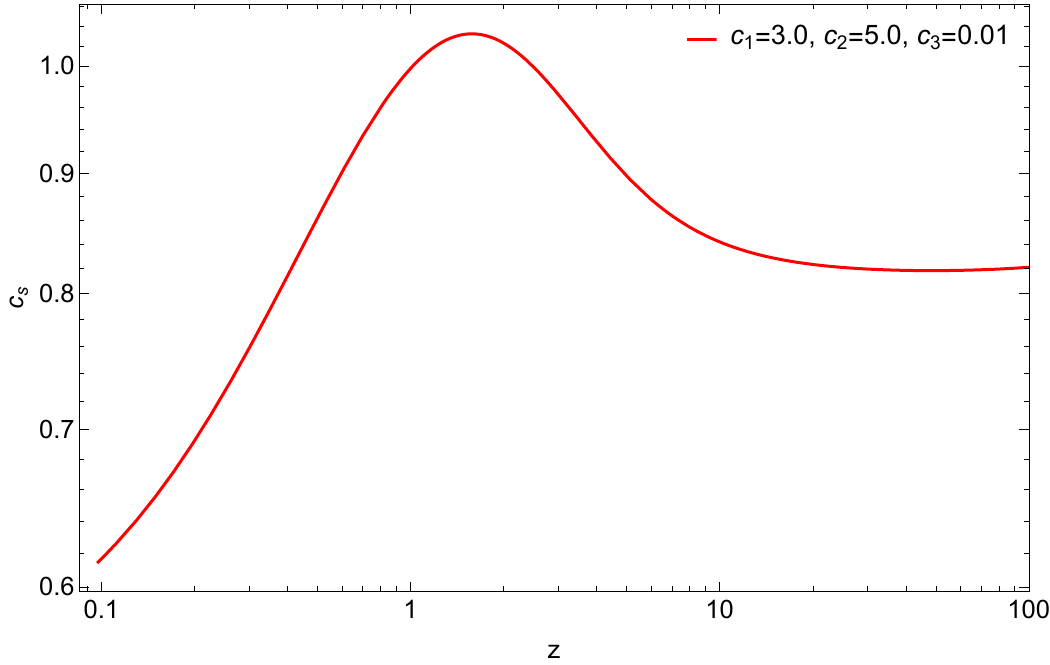}
    \includegraphics[width=0.48\textwidth]{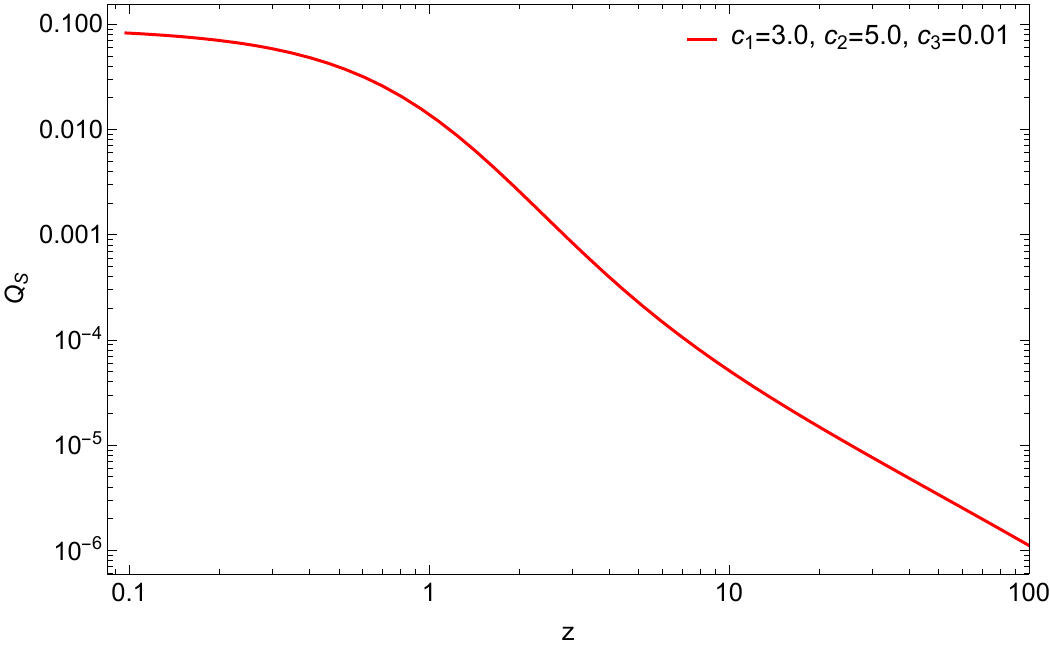}
    \caption{The evolution of the speed of scalar perturbations $c_s$ (left) and stability parameter $Q_s$ (right) for a specific choice of model parameters $c_1,c_2$ and $c_3$. The plots indicate that the model is free of gradient and ghost instabilities as specified by conditions in Eq. \ref{e: instability_check}.}
    \label{fig:allon_stability-parameters}
\end{figure}

For a consistent model building, it is necessary to ensure that there are no instabilities, as discussed in Sec. \ref{sec:horndeski_intro}. The necessary conditions to avoid such instabilities in a Horndeski setup are given by Eq. \ref{e: instability_check}. As shown in Fig. \ref{fig:allon_stability-parameters}, from the evolution of the speed of perturbations of scalar modes $c_S$ and the parameter $Q_S$, it is evident that our model is free of gradient and ghost instabilities. Moreover, as discussed earlier in Sec. \ref{sec:model}, for our setup given by Eq. \ref{e:setup}, the speed of tensor perturbations is always luminal thereby, we do not worry about tensor perturbations.

Finally, we comment upon the implications of such a model towards resolution of $\sigma_8$ (or $S_8$) tension as well, in reference to the points made by Heisenberg et al. in \cite{Heisenberg:2022lob}. The former reference discusses the necessary conditions in the background evolution of a dark energy model for simultaneous alleviation of $H_0$ and $\sigma_8$ tensions in a model-independent way. Our model meets the criteria suggested by \cite{Heisenberg:2022lob}, i.e., it exhibits a phantom crossing behaviour. Thereby, without going to the evolution of perturbations, we highlight that our model indeed shows the possibility of resolving the $\sigma_8$ tension as well. Due to the complicated structure of perturbations in a Horndeski setup, we leave the evolution of perturbations for future work.
\label{sec: features}

\subsection{Understanding the effects of the choice of $G_3$ and $G_4$ in our model}
\label{sec:appendix}
\begin{figure}
    \centering
    \includegraphics[width=0.48\textwidth]{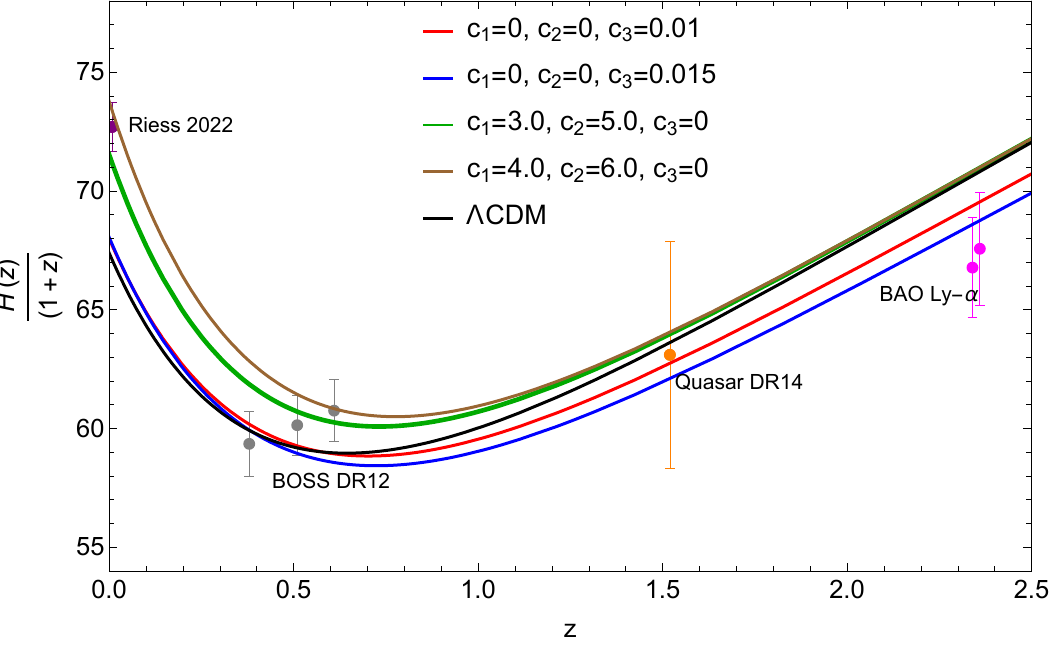}
    \includegraphics[width=0.48\textwidth]{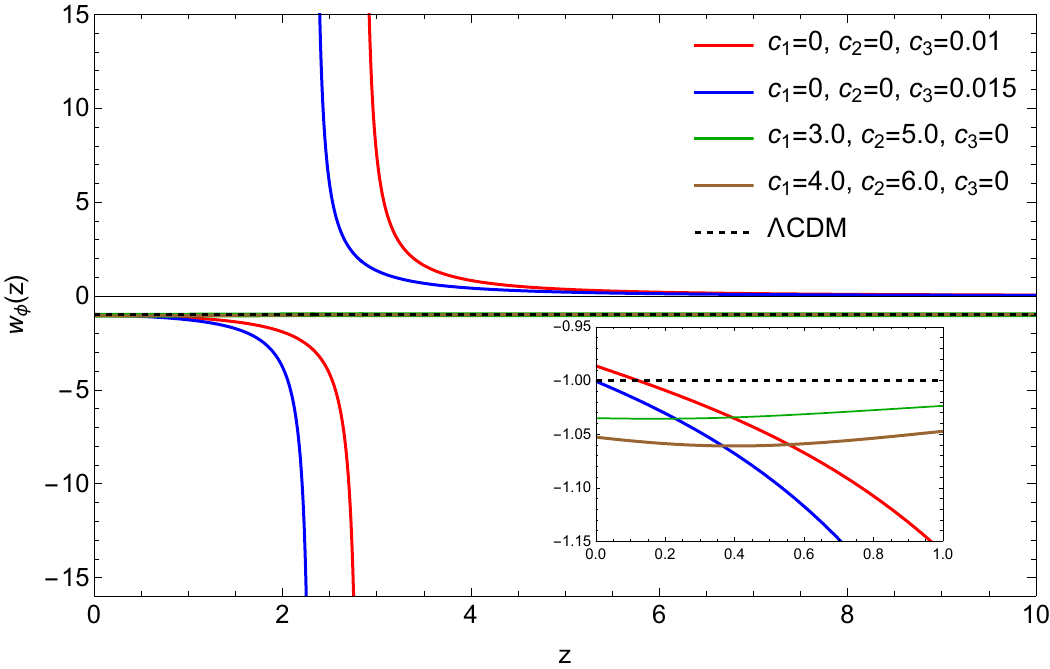}
    \caption{The evolution of $H(z)/(1+z)$ confronted with measurement of Hubble parameter from \cite{Riess:2021jrx}, BAO, and CC data \cite{Yu:2017iju, Ata:2017dya} (left) and the corresponding equation of state of dark energy scalar field (right) for the cases where either of $G_3$ and $G_4$ are turned on to see the effects of these individual terms on the background cosmological dynamics.}
    \label{fig:either_on}
\end{figure}

 As discussed earlier, the proposed dark energy model given by Eq. \ref{e:setup} involves non-zero $G_3$ and $G_4$ terms. From Eq. \ref{e: horndeski_lag}, one can see that a non-trivial $G_3$ invokes self-interactions of dark energy field while the choice of $G_4$ introduces non-minimal coupling to gravity. In this section, we will try to understand the individual effects of $G_3$ and $G_4$ on the dynamics of our model. As mentioned in Sec \ref{sec:model}, the role of $G_3$ is to lead to a phantom behaviour at low redshifts. The condition for a phantom behaviour is given by 
\begin{equation*}
    w_{\phi} = \frac{p_{\phi}}{\rho_{\phi}}<-1
\end{equation*}
For the case when $G_4$ is turned off, putting $c_3=0$ in Eqs. \ref{e:rho-ourmodel} and \ref{e:p-ourmodel}, the above condition resorts to,
\begin{equation*}
    \dot\phi^2(1-2c_1-c_2\Ddot{\phi}+3c_2 H \dot\phi)<0
\end{equation*}
which can be achieved with appropriate choices of $c_1$ (predominantly) and $c_2$. A non-zero $c_2$ is necessary to ensure the stability condition $Q_s>0$ is not violated in any regime. It is to be noted that with $c_3=0$, one cannot go into the $\rho_{\phi}<0$ regime as can be seen from Eq. \ref{e:rho-ourmodel}, since the terms $\mathcal{O}(\dot\phi^2) < V(\phi)$. Now, for the case when only the non-minimal coupling term $G_4$ is switched on, putting $c_1=c_2=0$, one gets
\begin{equation*}
    \rho_\phi=V(\phi)+\frac{1}{2}\dot\phi^2-6 c_3 \phi H^2-6 c_3 H \dot\phi
\end{equation*}
which can give rise to $\rho_{\phi}<0$ for a suitable choice of $c_3$ whenever $V(\phi)<6 c_3 \phi H^2$ in some regimes, where we ignore the terms $\mathcal{O}(\dot\phi)$ being very small. Since we are working with a linear potential, the above condition can be given as $V_0< 6 c_3 H^2$, or $V_0< 2 c_3 \rho_c$, where $\rho_c=3H^2$. This can easily be satisfied at high redshifts where dark energy density is really insignificant. We now show these results obtained from numerical solutions of the background equations Eqs. \ref{e:friedman-first}, \ref{e:friedman-second} and \ref{e:phi-ourmodel} for these two cases discussed above. The left panel in Fig. \ref{fig:either_on} shows the evolution of $H(z)/(1+z)$ for cases where either $G_3$ or $G_4$ is turned on. Note that we do not vary the scalar field mass (quantified by $V_0$) throughout the analysis. For the case with $G_3$ turned on (i.e. with $c_3=0$), one can obtain a higher value of $H_0$ arising due to the phantom behaviour of dark energy around the present epoch. This can be visualized from the green and brown plots in the evolution of $H(z)$ and $w_\phi(z)$ in Fig. \ref{fig:either_on}. But at the same time, the equation of state does not cross the phantom divide or become singular in any regime; thereby, such a setup may not address the BAO Ly-$\alpha$ measurements of $H(z)$ at high redshifts. In fact, it will not satisfy the condition of $\delta H(z)<0$ at high redshifts, as required for successful alleviation of the Hubble tension \cite{Heisenberg:2022lob,Adil:2023exv}. Such type of scenarios where the dark energy equation of state makes a sudden transition to phantom behaviour at very low redshifts,  so-called ``hockey stick" models, are not considered suitable (see for example \cite{Camarena:2021jlr}).
In the other case, when only $G_4$ term is turned on (by putting $c_1=c_2=0$), we do not obtain $H_0$ significantly higher than $\Lambda$CDM. This is also understood from the evolution of $w_\phi(z)$ in the red and blue curves in Fig. \ref{fig:either_on}, which rather behaves like a quintessence field at present redshift with $w_\phi>-1$. Interestingly, such a setup gives rise to a singularity in $w_\phi(z)$, which leads to a phantom crossing behaviour, the position of which can be controlled depending on the choice of parameter $c_3$. Because of this feature in $w_\phi$, arising due to negative dark energy at high redshifts, such setup can explain the $H(z)$ observations from BAO Ly-$\alpha$ measurements.  In conclusion, the phantom behaviour of the model can be attributed to the $G_3$ term (controlled by $c_1$ and $c_2$) while the negative dark energy at high redshifts arises due to the choice of $G_4$ (controlled by $c_3$) in our model given by Eq. \ref{e:setup}. The former effect gives rise to a large $H_0$ at present, explaining the observations from Supernovae \cite{Riess:2021jrx} while the latter gives the necessary phantom crossing condition to successfully resolve $H_0$ tension, in addition to explaining the measurements of $H(z)$ from BAO and SDSS at higher redshifts \cite{Ata:2017dya,Yu:2017iju}.
Hence, the combined characteristics drive the model towards a successful alleviation of the Hubble tension.

\subsection{Best fit constraints on model parameters}
In this section, we assess our model against the low redshift data to obtain constraints on the model parameters. This is achieved by employing a Markov Chain Monte Carlo (MCMC) analysis, which samples the parameter space of $c_1$, $c_2$ and $c_3$, to obtain best fit constraints on these parameters and thereby on the derived background quantities of interest (here $H_0$, $\Omega_M$) for different likelihoods. The data used in our analysis are as follows:
\begin{itemize}
    \item The SH0ES measurement of $H_0=73.3\pm 1.04$ km/s/Mpc \cite{Riess:2021jrx}, modelled with a Gaussian likelihood. 
    \item The distance moduli measurements of 1048 SNIa Pantheon sample in the redshift range $0.01<z<2.3$ \cite{Pan-STARRS1:2017jku}.
    \item The expansion rate measurements from BAO, compiled in Table \ref{tab: Table1}.
    \item The cosmic chronometer (CC) data on $H(z)$ as compiled in Table III of \cite{Gomez-Valent:2023uof} and the covariance matrix obtained\footnote {\url{https://gitlab.com/mmoresco/CCcovariance/-/tree/master?ref_type=heads}} following the method discussed in \cite{Moresco:2020fbm}.
\end{itemize}
For the sampling process, the three free model parameters are provided with uniform priors as summarised in Table \ref{tab: Table2}.
\begin{table}[ht]
\parbox{.5\linewidth}{
    \centering
    \begin{tabular}{|c|c|c|}
    \hline
    $z$ & $H(z)$ & Reference \\
    \hline
    0.38 & $81.9\pm 1.9$ & \cite{BOSS:2016wmc} \\
    0.51 & $90.8\pm 1.9$ & \cite{BOSS:2016wmc} \\
    0.61 & $97.8\pm 2.1$ & \cite{BOSS:2016wmc} \\
    1.52 & $159\pm 12$ & \cite{Zarrouk:2018vwy} \\
    2.34 & $223\pm 7$ & \cite{BOSS:2014hwf}\\
    2.36 & $227\pm 8$ & \cite{BOSS:2013igd}\\
    \hline
    \end{tabular}
    \caption{Baryon acoustic oscillation (BAO) expansion rate measurements from SDSS collaboration. The references of these measurements are provided in the table. \label{tab: Table1}}
}
\hspace{20pt}
\parbox{.5\linewidth}{
    \centering
    \begin{tabular}{|c|c|}
    \hline
    Parameters & Priors  \\
    \hline
    $c_1$ & [2.0,6.0]  \\
    $c_2$ & [3.0,10.0]  \\
    $c_3$ & [0.0,0.02]  \\
    \hline
    \end{tabular}
    \caption{The priors on model parameters used in our analysis. The prior range is the same for all the likelihoods used. \label{tab: Table2}}
}
\end{table}
\begin{figure}
    \centering
    \includegraphics[width=0.62\textwidth]{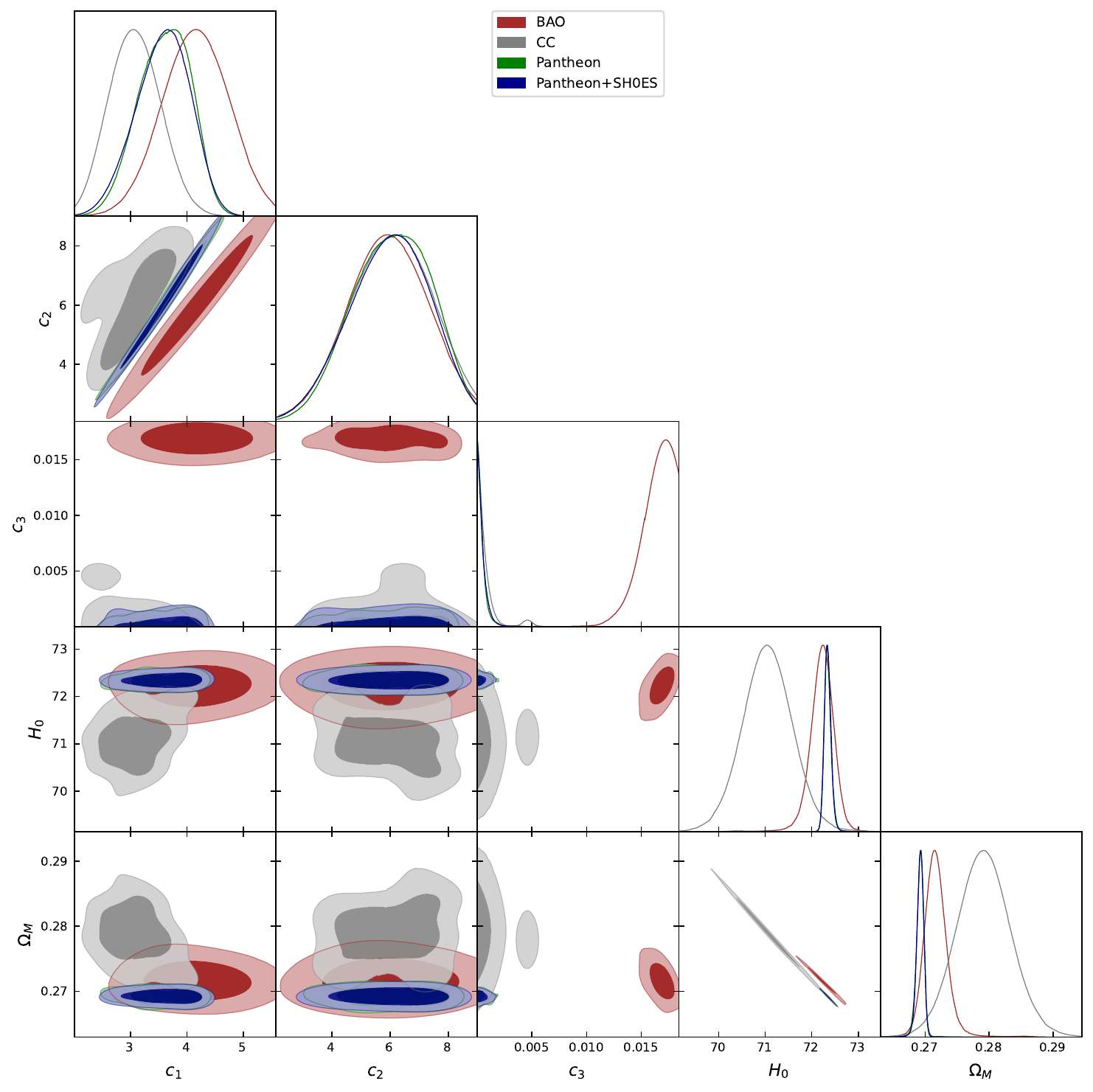}\\
    \includegraphics[width=0.62\textwidth]{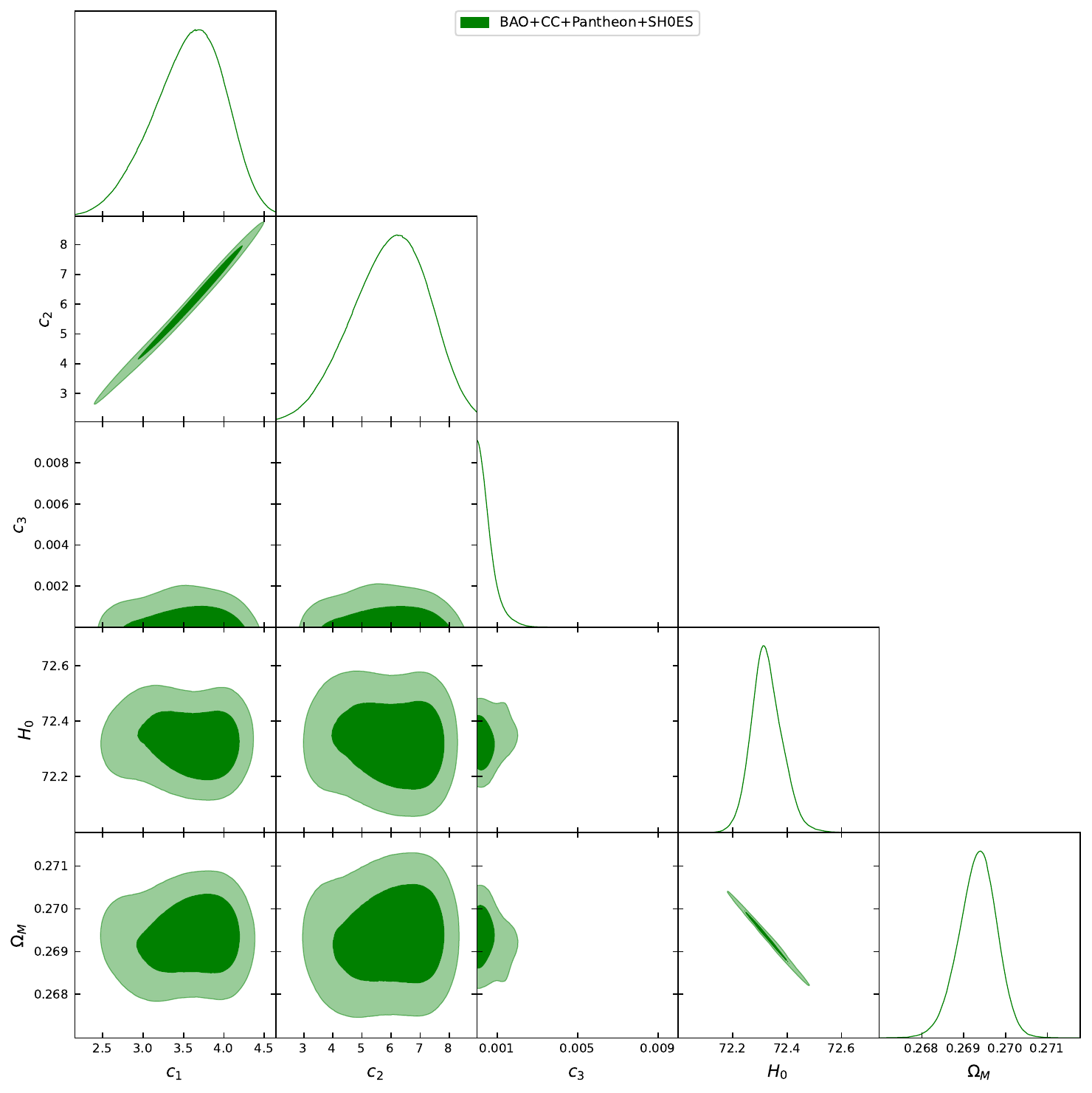}
    \caption{The posterior distribution of the model parameters $c_1,c_2,c_3$ and the derived background quantities $H_0$ and $\Omega_M$ from MCMC sampling for individual low redshift data (top panel) and for combined data (bottom panel).}
    \label{fig:posterior}
\end{figure}
In Fig \ref{fig:posterior} the results of the analysis are presented as posterior distributions of the model parameters and the derived background quantities for individual data (top panel) and for the combined data (bottom panel) obtained using the GetDist Python package \cite{Lewis:2019xzd}. We also tabulate the mean values and 1$\sigma$ constraints on these parameters obtained from the MCMC sampling in Table \ref{tab: Table3}.\par
For the case of individual likelihoods, the posteriors of parameters $c_1$ and $c_2$ overlap to a great extent for all the cases. Interestingly for the parameter $c_3$, which controls the strength of nonminimal coupling and causes the dark energy density to possess negative values at high redshifts, the constraints from BAO differ largely from those obtained for the other three data, namely CC, Pantheon and SH0ES. The BAO constraints seem to prefer a large value of $c_3$, thereby inclined to strongly favour the presence of negative dark energy at high redshifts and, thus, explain the anomalous Ly-$\alpha$  measurement of expansion rate, as also discussed in \cite{BOSS:2014hwf, Sahni:2014ooa}. But unlike BAO, the other data (CC, Pantheon and SH0ES) seem to be less sensitive to the nonminimal coupling parameter $c_3$, as evident from Table \ref{tab: Table3}, only providing an upper bound on it. This can also be understood, given that these data mostly constrain the expansion history at low redshifts ($z<2$), while the negative dark energy density feature is effective at somewhat larger redshifts. 
\par
A similar behaviour is seen in the joint likelihood analysis (with all data combined), where the parameter $c_3$ favours very small values with an upper bound as shown in Table \ref{tab: Table3}, indicating that the combined data does not constrain well the nonminimal coupling parameter. But any non-zero value of $c_3$, irrespective of its amplitude, implies the presence of negative dark energy. As discussed earlier, this incorporates interesting features like phantom crossing in the model, which is essential in order to successfully address $H_0$ and $\sigma_8$ tension without disturbing the CMB peaks \cite{Heisenberg:2022lob}. The other two parameters $c_1$ and $c_2$ are well-constrained with nearly Gaussian posteriors and similar mean values for different data. This indicates the preference for a phantom behaviour of dark energy at low redshifts (due to the non-zero $c_1$ and $c_2$), which leads to a large value of $H_0\sim72$ km/sec/Mpc, thereby reducing the tension with local measurements. However, in order to obtain the complete picture, one needs to include the CMB data as well to obtain parameter constraints, which is beyond the scope of the present work. Indeed, including the CMB data would help us impose more stringent constraints, especially on the nonminimal coupling parameter $c_3$, indicating whether or not the data favours the presence of negative dark energy at high redshifts.

\begin{table}[!]
\renewcommand{\arraystretch}{2.0}
    \begin{tabular}{ |p{2.0cm}|p{2.3cm}|p{2.3cm}|p{2.3cm}|p{2.3cm}|p{2.3cm}|}
     \hline
 Parameters & BAO & CC & Pantheon & \thead{Pantheon+\\SH0ES} & \thead{BAO+CC+\\Pantheon+\\SH0ES}\\
 \hline
$c_1$ & $4.50^{+0.61}_{-0.37}$   & $3.09^{+0.43}_{-0.48}$  & $3.61^{+0.52}_{-0.43}$& $3.58^{+0.52}_{-0.44}$  & $3.57^{+0.49}_{-0.39}$\\ 
$c_2$ &$6.70^{+1.0}_{-0.51}$ & $6.0\pm1.2$ & $6.1\pm1.3$ & $6.0\pm1.3$ & $6.0^{+1.5}_{-1.3}$\\ 
$c_3$ & $0.01714^{+0.00061}_{-0.00052}$ & $<0.000735$ & $<0.000410$ & $<0.000460$ & $<0.000558$\\
\hline
$H_0$ & $72.28\pm0.27$ & $71.05\pm0.46$ & $72.342^{+0.059}_{-0.079}$& $72.342^{+0.053}_{-0.069}$ & $72.325^{+0.050}_{-0.061}$\\  
$\Omega_M$ &$0.2712\pm0.0019$ &$0.2791\pm0.0036$ & $0.26921^{+0.00060}_{-0.00049}$ &$0.26921^{+0.00054}_{-0.00043}$& $0.26934^{+0.00047}_{-0.00042}$ \\
\hline
\end{tabular}
\vskip 10pt
    \caption{We present the best fit constraints on various model parameters after MCMC analysis with different low redshift data. The mean values of all the parameters along with their 1-$\sigma$ constraints are listed here.}
    \label{tab: Table3}
\end{table}

\section{Conclusions and Discussions}
\label{sec:conclusion}

The Hubble tension is certainly one of the most intriguing problems in modern cosmology and has recently received a lot of attention. Numerous attempts are ongoing as well as have been made in order to bridge the gap between the $\Lambda$CDM-predicted value of $H_0$ and the values obtained through the late-time (low redshift) observations. While the mismatch can certainly be attributed to the systematic errors associated with different direct or indirect (or both) measurements, the tension may also be a true reflection of reality, requiring exotic new physics and, perhaps, a dramatic revision to our current understanding of cosmic evolution on large scales.

 In this work, we have proposed a new dark energy model within the framework of Horndeski gravity as a plausible scenario towards the resolution of $H_0$ tension.
 The dark energy expansion is governed by a dynamical scalar field involving non-trivial self-interactions and non-minimal coupling, motivated by the Horndeski Lagrangian. The proposed model resorts to late-time modification approaches to resolve $H_0$ tension. 
 The two crucial characteristics of our model, which are negative dark energy density at high redshifts (giving rise to phantom crossing)  and a phantom behaviour around the present epoch,  make it an appealing candidate among various solutions to alleviate the Hubble tension. 
 As discussed in detail, both these conditions suffice for explaining the observed measurement of the Hubble parameter from independent
observation and at the same time, ensuring that the CMB measurements of the angular size of the sound horizon from Planck remain unaffected. 
We have further shown that our scenario remains free of various issues, such as gradient instability or superluminal propagation and, thus, is consistent from a physical model-building perspective. We have also briefly mentioned that our model shows the possibility of resolving the $\sigma_8$ tension as well due to the presence of a phantom crossing scenario \cite{Heisenberg:2022lob}. To check the feasibility of our model, we use low redshift data from Pantheon, SH0ES, BAO and CC, and employing an MCMC analysis, obtain the best fit constraints on model parameters. The analysis shows that the proposed dark energy model can support larger values of $H_0$($\sim 72$ km/sec/Mpc) due to the phantom behaviour of the dark energy field at low redshifts (due to self-coupling parameters $c_1$ and $c_2$), thereby resolving the $H_0$ tension. 

In fact, the next relevant step in this direction would be to study the evolution of linear density perturbations in such Horndeski models to see the effects on CMB and the matter power spectrum. 
We consider this as a future prospect to perform complete data analysis, including full CMB data, and explore the parameter space of such late Universe modifications arising within the Horndeski theory, along with examining the possibility of alleviating the $\sigma_8$ tension simultaneously. Finally, it would be very interesting to construct other consistent models within this framework, which can simultaneously alleviate the $H_0$ and $\sigma_8$ tensions without affecting the CMB measurements. We leave these fascinating possibilities for future work.

\section*{Acknowledgments}
BG would like to acknowledge financial support from the DST-INSPIRE Faculty Fellowship grant No. DST/INSPIRE/04/2020\\/
001534. RKJ acknowledges support from the Science and Engineering 
Research Board, Department of Science and Technology, Government of India, 
through the Core Research Grant No.~CRG/2018/002200 and the MATRICS grant~MTR/2022/000821.
RKJ also acknowledges financial support from the new faculty seed start-up 
grant of the Indian Institute of Science, Bengaluru, India, and the Infosys Foundation, Bengaluru, India through the Infosys Young Investigator Award. BG and YT thank Ujjwal Kumar Upadhyay for his useful comments.

\bibliography{references}
\end{document}